\documentclass[
    ,final            
  ]
  {aipproc}

\layoutstyle{8x11double}

\usepackage{psfig,graphics,graphicx}
\usepackage{color}
\usepackage{amssymb}
\usepackage{bookmath}

\begin{document}

\title{Knight Shift in the FFLO State of a Two-Dimensional D-Wave Superconductor}

\classification{75.60.-d, 76.60.Cq, 74.81.-g \hfill LA-UR-05-5255}

\author{Anton B. Vorontsov}{
  address={Department of Physics and Astronomy, Louisiana State University,
  Baton Rouge, LA 70803, USA}
}

\author{Matthias J. Graf}{
  address={Theoretical Division, Los Alamos National Laboratory,
  Los Alamos, NM 87545, USA}
}

\begin{abstract}
We report on the Fulde-Ferrell-Larkin-Ovchinnikov (FFLO) state in two-dimensional 
$d$-wave superconductors with magnetic field parallel to the superconducting planes. 
This state occurs at high magnetic field near the Pauli-Clogston limit and is a 
consequence of the competition between the pair condensation and Zeeman energy. 
We use the quasiclassical theory to self-consistently compute the spatially 
nonuniform order parameter.
Our self-consistent calculations show that the FFLO state of a $d$-wave order parameter 
breaks translational symmetry along preferred directions. The orientation of the 
nodes in real space is pinned by the nodes of the basis function in momentum space. 
Here, we present results for the Knight shift and discuss the implications for 
recent nuclear magnetic resonance measurements on CeCoIn$_5$.
\end{abstract}

\maketitle



The Fulde-Ferrell-Larkin-Ovchinnikov (FFLO) state of spin-singlet superconductors is the compromise
between the pairing condensate, favoring anti-parallel spin alignment, and the 
Zeeman effect, favoring parallel spin alignment along the field \cite{Fulde:1964,Larkin:1964}. 
This compromise leads to a spatially inhomogeneous state of ``normal'' and ``superconducting'' regions,
where the ``normal'' regions are defined by a spectrum of spin-polarized quasiparticles.

The FFLO phase of $d$-wave superconductors is modified by the anisotropy 
of the order parameter in momentum space compared to $s$-wave superconductors. The upper critical 
transition line, $B_{c2}(T)$, has a kink at low temperatures, $T^* \sim 0.06 \, T_c$, 
corresponding to the discontinuous change in the modulation of the order 
parameter \cite{Maki:1996,Shimahara:1997,Klein:2000,Vorontsov:2005}. 
Recent calculations of the spatial modulation of the order parameter in 2D near $B_{c2}$
predicted that the energetically favored state at low $T$ and high $B$ 
forms a ``square lattice'' instead of the 1D stripe order \cite{Shimahara:1998,Maki:2002}.

Here, we restrict our study to temperatures above this structural phase transition, $T > T^*$,
and address the quasiparticle response in the FFLO phase 
between the lower critical field $B_{c1}$ and the upper critical field $B_{c2}$. 
In addition, we assume that $\vB$ is parallel to the superconducting planes.  In this geometry 
the magnetic field affects the superconducting condensate only through the Zeeman coupling 
of the quasiparticle spin to the field.  Furthermore, we assume a cylindrical Fermi surface.


Within the quasiclassical theory of superconductivity we calculate self-consistently the 
order parameter $\Delta(\vR,\hat{\vp})$ and the quasiclassical Green's functions 
by solving Eilenberger's equation in a constant magnetic field $\vB$.
The Zeeman coupling of the quasiparticle spin with magnetic field enters through
$\mu \vB \cdot \vsigma$,
where $\sigma_i$ are Pauli spin matrices and $\mu = (g/2)\mu_B$ is 
the absolute value of the magnetic moment of a quasiparticle with negative charge $e$; 
$\mu_B=|e|/2mc$ is the Bohr magneton. Note that the $g$-factor is a free material 
parameter in this calculation.

From the solutions of Eilenberger's equation we can calculate 
measurable quantities like the free energy, quasiparticle local density of states 
\cite{Vorontsov:2005} and local magnetization $\vM(\vR)$.
Here, we consider spin-singlet order parameters that factorize into  
$\Delta(\vR,\hat{\vp})=\Delta(\vR)\,\cos 2\phi$,
with a spatially dependent amplitude, $\Delta(\vR)$, and an angular dependent
$d_{x^2-y^2}$-wave basis function.

The local magnetization is given by the paramagnetic response of the medium
and the spin-vector component of the quasiclassical Matsubara Green's function \cite{Alexander:1985}:
\be
\vM(\vR) = 2 \mu N_f 
\Big[ \mu\vB
 + T\, \sum_{\vare_n} \int \done{\hat{\vp}} \; \vg(\hat{\vp},\vR; \vare_n) 
\Big]
\,,
\ee
with the normal-state density of states $N_f$ at the Fermi level. 
The normal-state susceptibility, $\chi_N = 2 \mu^2 N_f$, is defined by
$M_N = \chi_N B$.

For comparison, we show in Fig.~1 the calculated temperature dependence of the magnetization in the 
uniform superconducting (USC) phase for three different values of $B$.
Increasing $B$ changes the $T$-dependence of the magnetization from linear to quadratic with 
a residual zero-temperature value due to the field induced shift in the spin-split density
of states of the gapless $d$-wave superconductor. This result is in agreement with
scaling arguments by Yang and Sondhi \cite{Yang:1998}.

In Fig.~2, we show temperature scans of the minimum, average and maximum local magnetization for the
stable FFLO phase, with spatial order-parameter modulation along the 
$\langle 110 \rangle$ direction, i.e., along the nodal direction of the gap function.
The Knight shift $K$ is proportional to the change in the local magnetic field at the nucleus, 
thus it is directly proportional to the local magnetization. Since $K$ is weighted by
the field distribution, the largest contribution comes from areas where the derivative 
of $\vM$ vanishes, which are at the minimum and maximum locations of $\vM$.
The calculated bifurcation between minimum and maximum local magnetization seen in Fig.~2 is in 
qualitative agreement with measurements of the Knight shift on CeCoIn$_5$ reported by
Kakuyanagi et al. \cite{NMR:2005}.

In Fig.~3, we show field scans of the local magnetization at $T / T_c = 0.1$
starting in the USC phase and into the FFLO $\langle 110 \rangle$ phase. 
It illustrates the nonlinear magnetic response of the quasiparticles due to an external field
and the continuous second order transition at the lower critical field $B_{c1}$, which also
is signaled by the appearance of a single domain wall.
This finding clearly contradicts the claim by Yang and Sondhi \cite{Yang:1998} about a first 
order transition at $B_{c1}$ between the USC and FFLO phase.

In addition, we calculated the spin-resolved local density of states in the FFLO state 
\cite{Vorontsov:2005} (not shown). We found that the characteristic Andreev bound states, due to 
the periodic sign change of the order parameter, are responsible for the excess spin polarization
of quasiparticles at the domain walls seen in Figs.~2 and 3.  Therefore, the Andreev bound states
should be clearly visible features in scanning tunneling spectroscopy measurements.


\begin{figure}
  \includegraphics[height=.28\textheight,angle=270]{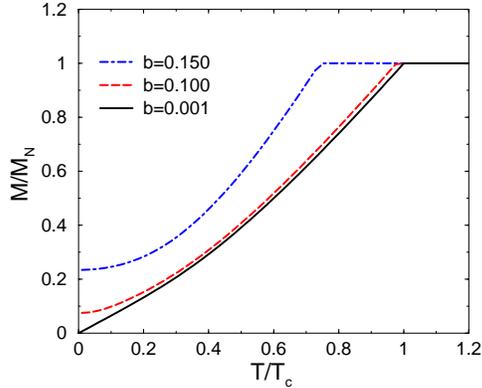}
  \caption{Magnetization $M$ of the spatially uniform superconducting (USC) phase normalized 
	by the normal-state magnetization $M_N$ for different magnetic fields $b=\mu B/2\pi T_c$.}
\end{figure}

\begin{figure}
  \includegraphics[height=.28\textheight,angle=270]{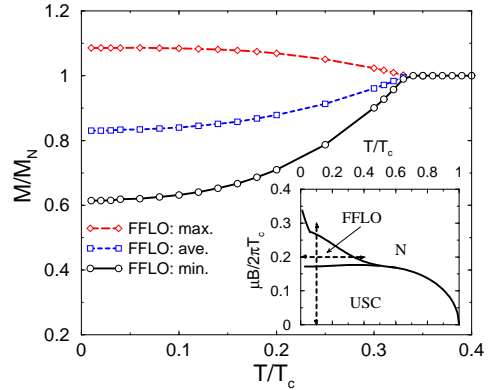}
  \caption{Temperature scan of the normalized local magnetization at fixed magnetic field $b=0.2$ in FFLO phase 
	for $\langle 110 \rangle$ orientation of spatial order-parameter modulation. Inset: Phase diagram
	with temperature and field scans across FFLO phase.}
\end{figure}

\begin{figure}
  \includegraphics[height=.28\textheight,angle=270]{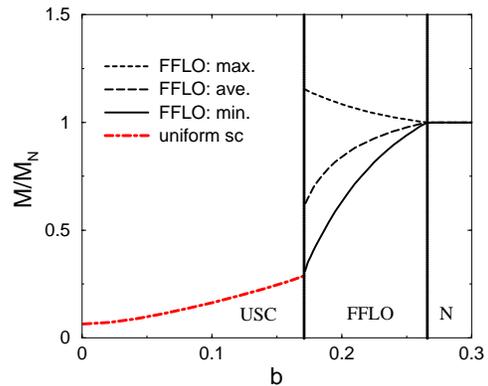}
  \caption{Field scan of the normalized local magnetization at fixed temperature $T/T_c=0.1$ across
	USC and FFLO phases for $\langle 110 \rangle$ orientation of order-parameter modulation.}
\end{figure}


\begin{theacknowledgments}
We thank N. J. Curro, R. Movshovich and V. F. Mitrovi\'c for helpful discussions
and A. V. Balatsky and J. A. Sauls for suggesting this problem.
This research is supported by the Department of Energy, under contract W-7405-ENG-36,
(MJG) and the LSU Board of Regents (ABV).
\end{theacknowledgments}



\bibliographystyle{aipprocl} 

\begin{thebibliography}{9}

\bibitem{Fulde:1964}
P. Fulde and R. Ferrell, \emph{Phys. Rev.} \textbf{135}, A550--A563 (1964).

\bibitem{Larkin:1964}
A. I. Larkin and Y. N. Ovchinnikov, \emph{Sov. Phys. JETP} \textbf{20}, 762--769 (1965).

\bibitem{Maki:1996}
K. Maki and H. Won, \emph{Czech. J. Phys.} \textbf{46}, 1035--1036 (1996).

\bibitem{Shimahara:1997}
H. Shimahara and D. Rainer, \emph{J. Phys. Soc. Japan} \textbf{66}, 3591--3599 (1997).

\bibitem{Klein:2000}
U. Klein, D. Rainer, and H. Shimahara, \emph{J. Low Temp. Phys.} \textbf{118}, 91--104 (2000).

\bibitem{Vorontsov:2005}
A. B. Vorontsov, J. A. Sauls, and M. J. Graf, \emph{cond-mat/0506257} (2005).

\bibitem{Shimahara:1998}
H. Shimahara, \emph{J. Phys. Soc. Japan} \textbf{67}, 736--739 (1998).

\bibitem{Maki:2002}
K. Maki and H. Won, \emph{Physica B} \textbf{322}, 315--317 (2002).

\bibitem{Alexander:1985}
J. A. X. Alexander, T. P. Orlando, D. Rainer,  and P. M. Tedrow, 
\emph{Phys. Rev. B} \textbf{31}, 5811--5825 (1985).

\bibitem{Yang:1998}
K. Yang and S. L. Sondhi, \emph{Phys. Rev. B} \textbf{57}, 8566 (1998).

\bibitem{NMR:2005}
K. Kakuyanagi et al., \emph{Phys. Rev. Lett.} \textbf{94}, 047602 (2005).

\end{thebibliography}

\end{document}